\documentclass[Conference,a4paper]{IEEEtran}

\IEEEoverridecommandlockouts

\usepackage{color}
\usepackage{graphicx}
\usepackage{amsmath}
\usepackage{amssymb}
\usepackage{algorithm}
\usepackage{algorithmic}
\usepackage{amsmath}
\usepackage{amsthm}
\usepackage{stfloats}
\usepackage{caption}
\usepackage{subfigure}
\usepackage{bm}
\usepackage{setspace}
\usepackage{flushend}

\newcommand{\be}{\begin{equation}}
\newcommand{\ee}{\end{equation}}

\newcommand{\non}{\nonumber}

\allowdisplaybreaks[4]

\title{Joint Beamforming and Reflection Design for RIS-assisted ISAC Systems
}
\author{\IEEEauthorblockN{Rang Liu$^{\dag}$, Ming Li$^{\dag}$, and A. Lee Swindlehurst$^{\ddag}$
\vspace{-0.0 cm} }\\
\IEEEauthorblockA{$^{\dag}$School of Information and Communication Engineering\\
Dalian University of Technology, Dalian, Liaoning 116024, China \\ E-mail: \texttt{liurang@mail.dlut.edu.cn, mli@dlut.edu.cn} } \\
\IEEEauthorblockA{$^{\ddag}$Center for Pervasive Communications and Computing\\ University of California, Irvine, CA 92697, USA\\ E-mail: \texttt{swindle@uci.edu} }

\vspace{0.2 cm}
\textit{(Invited Paper)}
}

\begin{document}

\maketitle

\pagestyle{empty}
\thispagestyle{empty}

\begin{abstract}
In this paper, we investigate the potential of employing reconfigurable intelligent surface (RIS) in integrated sensing and communication (ISAC) systems.
In particular, we consider an RIS-assisted ISAC system in which a multi-antenna base station (BS) simultaneously performs multi-user multi-input single-output (MU-MISO) communication and target detection.
We aim to jointly design the transmit beamforming and receive filter of the BS, and the reflection coefficients of the RIS to maximize the sum-rate of the communication users, while satisfying a worst-case radar output signal-to-noise ratio (SNR), the transmit power constraint, and the unit modulus property of the reflecting coefficients.
An efficient iterative algorithm based on fractional programming (FP), majorization-minimization (MM), and alternative direction method of multipliers (ADMM) is developed to solve the complicated non-convex problem.
Simulation results verify the advantage of the proposed RIS-assisted ISAC scheme and the effectiveness of the developed algorithm.
\end{abstract}

\section{Introduction}

While wireless communication and radar sensing have been separately developed for decades, integrated sensing and communication (ISAC) is recently arising as a promising technology for next-generation wireless networks.
ISAC not only allows communication and radar systems to share spectrum resources, but also enables a fully-shared platform transmitting unified waveforms to simultaneously perform communication and radar sensing functionalities, which significantly improves the spectral/energy/hardware efficiency  \cite{Liu-TCOM-2020}, \cite{Zhang-ICST-2022}.

Advanced signal processing techniques have been investigated for designing dual-functional transmit waveforms to achieve higher integration and coordination gains \cite{Zhang-JSTSP-2021}, \cite{Liu-TSP-2020}.
Although the transmit beamforming designs for multi-input multi-output (MIMO) systems significantly improve communication and radar sensing performance by exploiting spatial degrees of freedom (DoFs), deteriorated performance is still inevitable when encountering poor propagation conditions.
In such complex electromagnetic environments, the use of recently developed reconfigurable intelligent surface (RIS) technology can provide satisfactory performance by intelligently creating a favorable propagation environment \cite{Renzo-JSAC-2020}-\cite{Wu-TWC-2019}.

An RIS is generally a two-dimensional meta-surface consisting of many passive reflecting elements that can be independently tuned to establish favorable non-line-of-sight (NLoS) links between the transmitter and receivers.
Thus, additional DoFs are introduced by RIS for improving the system performance.
Inspired by this flexibility, the authors in \cite{Wang-TVT-2021a} and \cite{Wang-TVT-2021b} studied the employment of RIS to mitigate multi-user interference (MUI) and ensure sensing performance in terms of beampattern and Cram\'{e}r-Rao bound.
In \cite{Jiang-SJ-2021}, the signal-to-noise ratio (SNR) metric was utilized to evaluate the performance of an RIS-aided single-user system.
While most existing works simplify the system model and ignore the receive filter design for target detection, recent work \cite{Liu-2021} has presented comprehensive signal models and joint designs for the transmit waveform, receive filter and reflection coefficients.
However, the considered non-linear spatial-temporal transmit waveform and receive filter require more complicated hardware architectures and more complex algorithms.

Motivated by the above discussions, in this paper we investigate joint beamforming and reflection design for RIS-assisted ISAC systems, in which a multi-antenna base station (BS) delivers data to multiple single-antenna users and simultaneously performs target detection with the assistance of a single RIS.
The transmit beamformer and receive filter of the BS, and the RIS reflection coefficients are jointly optimized to maximize the communication sum-rate, as well as satisfy a minimum radar SNR constraint, the transmit power budget, and the unit modulus property of the reflecting elements.
To solve the resulting non-convex optimization problem, we employ fractional programming (FP), majorization-minimization (MM), and alternative direction method of multipliers (ADMM) methods to convert it into several tractable sub-problems and iteratively solve them.
Simulation studies illustrate the significant performance improvement introduced by the RIS and verify the effectiveness of the developed algorithm.

\section{System Model and Problem Formulation}

We consider an RIS-assisted ISAC system, where a colocated multi-antenna BS simultaneously performs multi-user communications and target detection with the assistance of an $N$-element RIS.
In particular, the BS, which is equipped with $M_\text{t}$ transmit antennas and $M_\text{r}$ receive antennas arranged as uniform linear arrays (ULAs) with half-wavelength spacing, simultaneously transmits data to $K$ single-antenna users and performs target detection, $M_\text{t}=M_\text{r}=M$ for simplicity.

The joint radar-communications signal that is transmitted in the $l$-th time slot is given by \cite{Liu-TSP-2020}
\be
\mathbf{x}[l] = \mathbf{W}_\text{c}\mathbf{s}_\text{c}[l] + \mathbf{W}_\text{r}\mathbf{s}_\text{r}[l]  = \mathbf{W}\mathbf{s}[l],
\ee
where $\mathbf{s}_\text{c}[l] \in\mathbb{C}^K$ contains the communication symbols for the $K$ users with $\mathbb{E}\{\mathbf{s}_\text{c}[l]\mathbf{s}_\text{c}^H[l]\}=\mathbf{I}_K$, $\mathbf{s}_\text{r}[l] \in\mathbb{C}^M$ includes $M$ individual radar waveforms with $\mathbb{E}\{\mathbf{s}_\text{r}[l]\mathbf{s}_\text{r}^H[l]\}=\mathbf{I}_M$, $\mathbb{E}\{\mathbf{s}_\text{c}[l]\mathbf{s}_\text{r}^H[l]\}=\mathbf{0}$, and  $\mathbf{W}_\text{c}\in\mathbb{C}^{M\times K}$ and $\mathbf{W}_\text{r}\in\mathbb{C}^{M\times M}$ denote the beamforming matrices for the  communication symbols and radar waveforms, respectively.
In addition, we define the beamforming matrix $\mathbf{W} \triangleq [\mathbf{W}_\text{c}~\mathbf{W}_\text{r}]$ and the symbol vector $\mathbf{s}[l]  \triangleq [\mathbf{s}_\text{c}^T[l] ~\mathbf{s}_\text{r}^T[l] ]^T$ for brevity.
Then, the received signal at the $k$-th user is expressed as
\be
y_k[l]  = (\mathbf{h}_{\text{d},k}^T + \mathbf{h}_{\text{r},k}^T\bm{\Phi}\mathbf{G})\mathbf{x}[l] + n_k[l],
\ee
where $\mathbf{h}_{\text{d},k}\in\mathbb{C}^M$, $\mathbf{h}_{\text{r},k}\in\mathbb{C}^N$, and $\mathbf{G}\in\mathbb{C}^{N\times M}$ denote the channels between the BS and the $k$-th user, between the RIS and the $k$-th user, and between the BS and the RIS, respectively.
The channels $\mathbf{h}_{\text{d},k}$ and $\mathbf{h}_{\text{r},k},~\forall k$, follow Rayleigh fading and the channel $\mathbf{G}$ is LoS.
The reflection matrix is defined as $\bm{\Phi}\triangleq\text{diag}\{\bm{\phi}\}$, where $\bm{\phi}\triangleq[\phi_1,\ldots,\phi_N]^T$ is the vector of reflection coefficients satisfying $|\phi_n|=1,~\forall n$.
The scalar $n_k[l]\sim\mathcal{CN}(0,\sigma_k^2)$ is additive white Gaussian noise (AWGN) at the $k$-th user.
Thus, the signal-to-interference-plus-noise ratio (SINR) of the $k$-th user can be calculated as\vspace{-0.1 cm}
\be
\text{SINR}_k = \frac{|\mathbf{h}^T_k(\bm{\phi})\mathbf{w}_k|^2}
{\sum_{j\neq k}^{K+M}|\mathbf{h}^T_k(\bm{\phi})\mathbf{w}_j|^2+\sigma_k^2},
\ee
where for conciseness we define $\mathbf{h}_k(\bm{\phi}) \triangleq \mathbf{h}_{\text{d},k} + \mathbf{G}^T\bm{\Phi}\mathbf{h}_{\text{r},k}$ as the composite channel between the BS and the $k$-th user, and $\mathbf{w}_j$ as the $j$-th column of $\mathbf{W}$, i.e., $\mathbf{W} = [\mathbf{w}_1,\ldots,\mathbf{w}_{K+M}]$.

Meanwhile, the echo signals collected by the BS receive array can be expressed as
\be
\mathbf{y}_\text{r}[l]  = \alpha_\text{t}(\mathbf{h}_{\text{d},\text{t}} + \mathbf{G}^T\bm{\Phi}\mathbf{h}_{\text{r},\text{t}})(\mathbf{h}^T_{\text{d},\text{t}} + \mathbf{h}^T_{\text{r},\text{t}}\bm{\Phi}\mathbf{G})\mathbf{W}\mathbf{s}[l] + \mathbf{n}_\text{r}[l],
\ee
where $\alpha_\text{t}$ is the complex target amplitude with $\mathbb{E}\{|\alpha_\text{t}|^2\}=\sigma_\text{t}^2$, $\mathbf{h}_{\text{d},\text{t}}\in\mathbb{C}^M$ and $\mathbf{h}_{\text{r},\text{t}}\in\mathbb{C}^N$ respectively represent the channels between the BS/RIS and the target, and $\mathbf{n}_\text{r}[l]\sim\mathcal{CN}(\mathbf{0},\sigma_\text{r}^2\mathbf{I}_M)$ is AWGN.
It is noted that the BS/RIS-target links are LoS and the angle of arrival/departure (AoA/AoD) of interest is known a priori.
The received signals over $L$ samples after the matched-filtering can be written as\vspace{-0.1 cm}
\be
\mathbf{Y}_\text{r} = \alpha_\text{t}\mathbf{H}_\text{t}(\bm{\phi})\mathbf{WSS}^H + \mathbf{N}_\text{r}\mathbf{S}^H,
\ee
where we define the equivalent channel for the target return as
$\mathbf{H}_\text{t}(\bm{\phi})\triangleq (\mathbf{h}_{\text{d},\text{t}} + \mathbf{G}^T\bm{\Phi}\mathbf{h}_{\text{r},\text{t}})(\mathbf{h}^T_{\text{d},\text{t}} + \mathbf{h}^T_{\text{r},\text{t}}\bm{\Phi}\mathbf{G})$, and the symbol/noise matrix over the $L$ samples as $\mathbf{S} \triangleq [\mathbf{s}[1],\ldots,\mathbf{s}[L]]$ and $\mathbf{N}_\text{r} \triangleq [\mathbf{n}_\text{r}[1],\ldots,\mathbf{n}_\text{r}[L]]$, respectively.
Defining $\widetilde{\mathbf{y}}_\text{r} \triangleq \text{vec}\{\mathbf{Y}_\text{r}\}$, $\mathbf{w} \triangleq \text{vec}\{\mathbf{W}\}$, and $\widetilde{\mathbf{n}}_\text{r} \triangleq \text{vec}\{\mathbf{N}_\text{r}\mathbf{S}^H\}$, the vectorized received signals can be expressed as\vspace{-0.1 cm}
\be
\widetilde{\mathbf{y}}_\text{r}  = \alpha_\text{t}(\mathbf{SS}^H\otimes\mathbf{H}_\text{t}(\bm{\phi}))\mathbf{w}+ \widetilde{\mathbf{n}}_\text{r}.
\ee

In order to achieve satisfactory target detection performance, a receive filter/beamformer $\mathbf{u}\in\mathbb{C}^{M\times (K+M)}$ is then applied to process $\widetilde{\mathbf{y}}_\text{r}$ and yields\vspace{-0.1 cm}
\be
\mathbf{u}^H\widetilde{\mathbf{y}}_\text{r}  = \alpha_\text{t}\mathbf{u}^H(\mathbf{SS}^H\otimes\mathbf{H}_\text{t}(\bm{\phi}))\mathbf{w} + \mathbf{u}^H\widetilde{\mathbf{n}}_\text{r}.
\ee
Therefore, the radar SNR for target detection is formulated as\vspace{-0.1 cm}
\be\label{eq:SINRt org}
\text{SNR}_\text{t} = \frac{\sigma_\text{t}^2\mathbb{E}\big\{|\mathbf{u}^H(\mathbf{SS}^H\otimes\mathbf{H}_\text{t}(\bm{\phi}))\mathbf{w}|^2\big\}}
{L\sigma_\text{r}^2\mathbf{u}^H\mathbf{u}}.
\ee
Since the numerator in (\ref{eq:SINRt org}) is complicated and difficult for optimization, we utilize Jensen's inequality, i.e., $\mathbb{E}\{f(x)\} \geq f(\mathbb{E}\{x\})$, and the fact that $\mathbb{E}\{\mathbf{SS}^H\} = L\mathbf{I}_{K+M}$ to obtain the following lower bound for the SNR:\vspace{-0.1 cm}
\be\label{eq:SINRt}
\text{SNR}_\text{t} \geq \frac{L\sigma_\text{t}^2|\mathbf{u}^H(\mathbf{I}_{K+M}\otimes\mathbf{H}_\text{t}(\bm{\phi}))\mathbf{w}|^2}
{\sigma_\text{r}^2\mathbf{u}^H\mathbf{u}},
\ee
which represents the achieved SNR in the worst case.

In this paper, we aim to jointly optimize the transmit beamforming $\mathbf{W}$, the receive filter $\mathbf{u}$, and the reflecting coefficients $\bm{\phi}$ to maximize the achievable sum-rate for multi-user communications, as well as satisfy the worst-case radar SNR $\Gamma_\text{t}$, the transmit power budget $P$, and the unit modulus property of the reflecting coefficients.
Therefore, the optimization problem is formulated as\vspace{-0.15 cm}
\begin{subequations}\label{eq:original problem}\begin{align}
&\underset{\mathbf{W},\mathbf{u},\bm{\phi}}\max~~\sum_{k=1}^K\log_2(1+\text{SINR}_k)\\
&\quad\text{s.t.}\quad~\frac{L\sigma_\text{t}^2|\mathbf{u}^H(\mathbf{I}_{K+M}\otimes\mathbf{H}_\text{t}(\bm{\phi}))\mathbf{w}|^2}
{\sigma_\text{r}^2\mathbf{u}^H\mathbf{u}} \geq \Gamma_\text{t}, \\
&\quad\quad\quad~~\|\mathbf{W}\|_F^2 \leq P,\\
&\quad\quad\quad~~|\phi_n| = 1,~~\forall n.
\end{align}
\end{subequations}
It is obvious that the non-convex problem (\ref{eq:original problem}) is very difficult to solve due to the complicated objective function (\ref{eq:original problem}a) with $\log(\cdot)$ and fractional terms, the coupled variables in both the objective function (\ref{eq:original problem}a) and the radar SNR constraint (\ref{eq:original problem}b), and the unit modulus constraint (\ref{eq:original problem}d).
In order to tackle these difficulties, in the next section we propose to utilize FP, MM, and ADMM methods to convert problem (\ref{eq:original problem}) into several tractable sub-problems and iteratively solve them.

\vspace{-0.1 cm}
\section{Joint Beamforming and Reflection Design}

\subsection{FP-based Transformation}

We start by converting the complicated objective function (\ref{eq:original problem}a) into a more favorable polynomial expression based on FP.
As derived in \cite{Shen-TSP-2018}, by employing the Lagrangian dual reformulation and introducing an auxiliary variable $\mathbf{r}\triangleq[r_1,\ldots,r_K]^T$, the objective (\ref{eq:original problem}a) can be transformed into
\be\label{eq:after FP1}
\sum_{k=1}^K\!\log_2(1+r_k) - \sum_{k=1}^K\!r_k + \sum_{k=1}^K\!\frac{(1+r_k)|\mathbf{h}^T_k(\bm{\phi})\mathbf{w}_k|^2}
{\sum_{j=1}^{K+M}\!|\mathbf{h}^T_k(\bm{\phi})\mathbf{w}_j|^2\!+\!\sigma_k^2},
\ee
in which the variables $\mathbf{w}$ and $\bm{\phi}$ are taken out of the $\log(\cdot)$ function and coupled in the third fractional term.
Then, expanding the quadratic terms and introducing an auxiliary variable $\mathbf{c}\triangleq[c_1,\ldots,c_K]^T$, (\ref{eq:after FP1}) can be further converted into
\be\label{eq:new obj}\begin{aligned}
&f(\mathbf{w},\bm{\phi},\mathbf{r},\mathbf{c}) \triangleq  \sum_{k=1}^K\log_2(1+r_k) - \sum_{k=1}^K r_k -\sum_{k=1}^K|c_k|^2\sigma_k^2 \\
& + \sum_{k=1}^K\!2\sqrt{1\!+\!r_k}\Re\{c_k^*\mathbf{h}^T_k(\bm{\phi})\mathbf{w}_k\} -\!\sum_{k=1}^K\!|c_k|^2\!\!\sum_{j=1}^{K+M}|\mathbf{h}^T_k(\bm{\phi})\mathbf{w}_j|^2.
\end{aligned}\ee

To facilitate the algorithm development, we attempt to re-arrange the new objective function (\ref{eq:new obj}) into explicit and compact forms with respect to $\mathbf{w}$ and $\bm{\phi}$, respectively.
By stacking the vectors $\mathbf{w}_j,~\forall j$, into $\mathbf{w}$ and applying $\mathbf{h}^T_k(\bm{\phi})\mathbf{w}_j = \mathbf{h}_{\text{d},k}^T\mathbf{w}_j + \mathbf{h}_{\text{r},k}^T\text{diag}\{\mathbf{G}\mathbf{w}_j\}\bm{\phi}$, equivalent expressions for $f(\mathbf{w},\bm{\phi},\mathbf{r},\mathbf{c})$ can be obtained as
\begin{subequations}\begin{align}
f(\mathbf{w},\bm{\phi},\mathbf{r},\mathbf{c}) &= \Re\{\mathbf{a}^H\mathbf{w}\}-\|\mathbf{B}\mathbf{w}\|^2 + \varepsilon_1\\
& = \Re\{\mathbf{g}^H\bm{\phi}\} - \bm{\phi}^H\mathbf{D}\bm{\phi} + \varepsilon_2,
\end{align}\end{subequations}
where we define
\begin{small}\begin{align}
\mathbf{a} &\triangleq [2\sqrt{1\!+\!r_k}c_k^*\mathbf{h}^T_k(\bm{\phi}),\ldots,
2\sqrt{1\!+\!r_K}c_K^*\mathbf{h}^T_K(\bm{\phi}),\mathbf{0}^T]^H,\non\\
\mathbf{B} &\triangleq [\mathbf{b}_{1,1} \ldots, \mathbf{b}_{K,K+M}]^T,\quad
\mathbf{b}_{k,j} \triangleq |c_k|\mathbf{T}_j^T\mathbf{h}_k(\bm{\phi}),\non\\
\varepsilon_1 &\triangleq \sum_{k=1}^K\log_2(1+r_k) - \sum_{k=1}^K r_k-\sum_{k=1}^K|c_k|^2\sigma_k^2,\non\\
\mathbf{g} &\triangleq 2\sum_{k=1}^K\sqrt{1+r_k}c_k\text{diag}\{\mathbf{w}_k^H\mathbf{G}^H\}\mathbf{h}_{\text{r},k}^* \non\\
&\quad - 2\sum_{k=1}^K|c_k|^2\sum_{j=1}^{K+M}\text{diag}\{\mathbf{w}_j^H\mathbf{G}^H\}\mathbf{h}_{\text{r},k}^*\mathbf{h}^T_{\text{d},k}\mathbf{w}_j,\non\\
\mathbf{D} &\triangleq \sum_{k=1}^K\!|c_k|^2\!\sum_{j=1}^{K+M}\text{diag}\{\mathbf{w}_j^H\mathbf{G}^H\}
\mathbf{h}_{\text{r},k}^*\mathbf{h}^T_{\text{r},k}\text{diag}\{\mathbf{G}\mathbf{w}_j\},\non\\
\varepsilon_2 &\triangleq \varepsilon_1 \!+\! \sum_{k=1}^K\!\big[2\sqrt{1\!+\!r_k}\Re\{c_k^*\mathbf{h}^T_{\text{d},k}\mathbf{w}_k\}\!-\!  |c_k|^2\!\!\sum_{j=1}^{K+M}\!\!|\mathbf{h}^T_{\text{d},k}\mathbf{w}_j|^2\big],\non
\end{align}\end{small}
\vspace{-0.2 cm}

\noindent and $\mathbf{T}_j\in\mathbb{C}^{M\times M(K+M)}$ as a permutation matrix to extract $\mathbf{w}_j$ from $\mathbf{w}$, i.e., $\mathbf{w}_j = \mathbf{T}_j\mathbf{w}$.
Now, we can clearly see that the re-formulated objective $f(\mathbf{w},\bm{\phi},\mathbf{r},\mathbf{c})$ is a conditionally concave function with respect to each variable given the others, which allows us to iteratively solve for each variable as shown below.

\vspace{-0.2 cm}
\subsection{Block Update}

\subsubsection{Update $\mathbf{r}$ and $\mathbf{c}$}

Given other variables, the optimization for the auxiliary variable $\mathbf{r}$ is an unconstrained convex problem, whose optimal solution can be easily obtained by setting $\frac{\partial f}{\partial\mathbf{r}} = \mathbf{0}$.
The optimal $r_k^\star$ is calculated as
\be\label{eq:update rk}
r_k^\star = \frac{|\mathbf{h}^T_k(\bm{\phi})\mathbf{w}_k|^2}
{\sum_{j\neq k}^{K+M}|\mathbf{h}^T_k(\bm{\phi})\mathbf{w}_j|^2+\sigma_k^2},~\forall k.
\ee
Similarly, the optimal $c_k^\star$ is obtained by setting $\frac{\partial f}{\partial c_k} = 0$ as
\be\label{eq:update ck}
c_k^\star = \frac{\sqrt{1+r_k}\mathbf{h}^T_k(\bm{\phi})\mathbf{w}_k}{\sum_{j=1}^{K+M}|\mathbf{h}^T_k(\bm{\phi})\mathbf{w}_j|^2
+\sigma_k^2},~\forall k.
\ee

\subsubsection{Update $\mathbf{u}$}

Finding $\mathbf{u}$ with the other parameters fixed leads to a feasibility check problem without an explicit objective.
In order to accelerate convergence and leave more DoFs for sum-rate maximization in the next iteration, we propose to maximize the SNR lower bound for updating $\mathbf{u}$.
Thus, the optimization problem is formulated as
\be\label{eq:u problem}
\underset{\mathbf{u}}\max~~\frac{L\sigma^2_\text{t}|\mathbf{u}^H(\mathbf{I}_{K+M}\otimes\mathbf{H}_\text{t}(\bm{\phi}))\mathbf{w}|^2}
{\sigma_\text{r}^2\mathbf{u}^H\mathbf{u}},
\ee
which is a typical Rayleigh quotient with the optimal solution
\be\label{eq:update u}
\mathbf{u}^\star = \frac{(\mathbf{I}_{K+M}\otimes\mathbf{H}_\text{t}(\bm{\phi}))\mathbf{w}}
{\mathbf{w}^H(\mathbf{I}_{K+M}\otimes\mathbf{H}^H_\text{t}(\bm{\phi})\mathbf{H}_\text{t}(\bm{\phi}))\mathbf{w}}.
\ee

Moreover, we see that $e^{\jmath\theta}\mathbf{u}^\star$ is also an optimal solution to (\ref{eq:u problem}) for an arbitrary angle $\theta$, since the phase of the output $\mathbf{u}^H\widetilde{\mathbf{y}}_\text{r}$ does not change the achieved SNR.
Inspired by this finding, after obtaining $\mathbf{u}$ we can restrict the term $\mathbf{u}^H(\mathbf{I}_{K+M}\otimes\mathbf{H}_\text{t}(\bm{\phi}))\mathbf{w}$ to be a non-negative real value, and thus re-formulate the radar output SNR constraint (\ref{eq:original problem}b) as
\be
\Re\{\mathbf{u}^H(\mathbf{I}_{K+M}\otimes\mathbf{H}_\text{t}(\bm{\phi}))\mathbf{w}\} \geq \varepsilon_3,
\ee
where for brevity we define $\varepsilon_3 \triangleq \sqrt{\Gamma_\text{t}\sigma_\text{r}^2\mathbf{u}^H\mathbf{u}/(L\sigma^2_\text{t})}$.

\newcounter{TempEqCnt}
\setcounter{TempEqCnt}{\value{equation}}
\setcounter{equation}{21}
\begin{figure*}[!t]
\begin{subequations}\label{eq:reformulate1}
\begin{align}
&(\mathbf{I}\otimes\mathbf{H}_\text{t}(\bm{\phi}))\mathbf{w} \non \\
&=\big(\mathbf{I}\!\otimes\!\mathbf{h}_{\text{d},\text{t}}\mathbf{h}_{\text{d},\text{t}}^T\big)\mathbf{w} +
\big(\mathbf{I}\!\otimes\!\mathbf{G}^T\!\text{diag}\{\mathbf{h}_{\text{r},\text{t}}\}\bm{\phi}\mathbf{h}_{\text{d},\text{t}}^T\big)\mathbf{w} + \big(\mathbf{I}\!\otimes\!\mathbf{h}_{\text{d},\text{t}}\bm{\phi}^T\text{diag}\{\mathbf{h}_{\text{r},\text{t}}\}\mathbf{G}\big)\mathbf{w}
+ \big(\mathbf{I}\!\otimes\!\mathbf{G}^T\text{diag}\{\mathbf{h}_{\text{r},\text{t}}\}\bm{\phi}\bm{\phi}^T\text{diag}\{\mathbf{h}_{\text{r},\text{t}}\}\mathbf{G}\big)\mathbf{w}\\
& = \big(\mathbf{I}\!\otimes\!\mathbf{h}_{\text{d},\text{t}}\mathbf{h}_{\text{d},\text{t}}^T\big)\mathbf{w} + \text{vec}\big\{\mathbf{G}^T\text{diag}\{\mathbf{h}_{\text{r},\text{t}}\}\bm{\phi}\mathbf{h}_{\text{d},\text{t}}^T\mathbf{W}
+ \mathbf{h}_{\text{d},\text{t}}\bm{\phi}^T\text{diag}\{\mathbf{h}_{\text{r},\text{t}}\}\mathbf{G}\mathbf{W}
+ \mathbf{G}^T\text{diag}\{\mathbf{h}_{\text{r},\text{t}}\}\bm{\phi}\bm{\phi}^T\text{diag}\{\mathbf{h}_{\text{r},\text{t}}\}\mathbf{G}\mathbf{W}\big\}
\\& = \big(\mathbf{I}\!\otimes\!\mathbf{h}_{\text{d},\text{t}}\mathbf{h}_{\text{d},\text{t}}^T\big)\mathbf{w} +
\big(\!\underbrace{\mathbf{W}^T\mathbf{h}_{\text{d},\text{t}}\!\otimes\!\mathbf{G}^T\!\text{diag}\{\mathbf{h}_{\text{r},\text{t}}\}
\!+\!\mathbf{W}^T\mathbf{G}^T\!\text{diag}\{\mathbf{h}_{\text{r},\text{t}}\}\!\otimes\!\mathbf{h}_{\text{d},\text{t}}}_{\mathbf{F}}\!\big)\bm{\phi}
+ \big(\!\underbrace{\mathbf{W}^T\mathbf{G}^T\!\text{diag}\{\mathbf{h}_{\text{r},\text{t}}\}\!\otimes\!\mathbf{G}^T\!\text{diag}\{\mathbf{h}_{\text{r},\text{t}}\}}_{\mathbf{L}}\!\big)\text{vec}\{\bm{\phi}\bm{\phi}^T\}.
\end{align}\end{subequations}
\vspace{-1 cm}

\rule[-12pt]{18.5 cm}{0.05em}
\end{figure*}
\setcounter{equation}{\value{TempEqCnt}}

\subsubsection{Update $\mathbf{w}$}

With fixed $\mathbf{r}$, $\mathbf{c}$, $\mathbf{u}$, and $\bm{\phi}$, the optimization for the transmit beamforming $\mathbf{w}$ can be formulated as
\begin{subequations}\label{eq:solve W}
\begin{align}
&\underset{\mathbf{w}}\min~~\|\mathbf{B}\mathbf{w}\|^2 -\Re\{\mathbf{a}^H\mathbf{w}\}\\
&~\text{s.t.}~~\Re\{\mathbf{u}^H(\mathbf{I}_{K+M}\otimes\mathbf{H}_\text{t}(\bm{\phi}))\mathbf{w}\} \geq \varepsilon_3,\\
&\quad\quad~\|\mathbf{w}\|^2 \leq P.
\end{align}
\end{subequations}
Obviously, this is a simple convex problem that can be readily solved by various well-developed algorithms or toolboxes.

\subsubsection{Update $\bm{\phi}$}

Given $\mathbf{r}$, $\mathbf{c}$, $\mathbf{u}$, and $\mathbf{w}$, the optimization for the reflection coefficients $\bm{\phi}$ is formulated as
\begin{subequations}\label{eq:solve phi org}
\begin{align}
&\underset{\bm{\phi}}\min~~\bm{\phi}^H\mathbf{D}\bm{\phi}-\Re\{\mathbf{g}^H\bm{\phi}\}\\
&~\text{s.t.}~~\Re\{\mathbf{u}^H(\mathbf{I}_{K+M}\otimes\mathbf{H}_\text{t}(\bm{\phi}))\mathbf{w}\} \geq \varepsilon_3,\\
&\quad\quad~|\phi_n| = 1,~~\forall n,
\end{align}
\end{subequations}
which cannot be directly solved due to the implicit function with respect to $\bm{\phi}$ in constraint (\ref{eq:solve phi org}b) and the non-convex unit modulus constraint (\ref{eq:solve phi org}c).

We first propose to handle constraint (\ref{eq:solve phi org}b) by re-arranging its left-hand side as an explicit expression with respect to $\bm{\phi}$ and then employing the MM method to find a favorable surrogate function for it.
Recall that $\mathbf{H}_\text{t}(\bm{\phi})\triangleq(\mathbf{h}_{\text{d},\text{t}} + \mathbf{G}^T\bm{\Phi}\mathbf{h}_{\text{r},\text{t}})(\mathbf{h}^T_{\text{d},\text{t}} + \mathbf{h}^T_{\text{r},\text{t}}\bm{\Phi}\mathbf{G}) = \mathbf{h}_{\text{d},\text{t}}\mathbf{h}^T_{\text{d},\text{t}} + \mathbf{G}^T\bm{\Phi}\mathbf{h}_{\text{r},\text{t}}\mathbf{h}^T_{\text{d},\text{t}} + \mathbf{h}_{\text{d},\text{t}}\mathbf{h}^T_{\text{r},\text{t}}\bm{\Phi}\mathbf{G} + \mathbf{G}^T\bm{\Phi}\mathbf{h}_{\text{r},\text{t}}\mathbf{h}^T_{\text{r},\text{t}}\bm{\Phi}\mathbf{G}$.
By employing the transformations $\bm{\Phi}\mathbf{h}_{\text{r},\text{t}} = \text{diag}\{\mathbf{h}_{\text{r},\text{t}}\}\bm{\phi}$ and $\text{vec}\{\mathbf{ABC}\}=(\mathbf{C}^T\otimes \mathbf{A})\text{vec}\{\mathbf{B}\}$, the term $(\mathbf{I}\otimes\mathbf{H}_\text{t}(\bm{\phi}))\mathbf{w}$ can be equivalently transformed into (\ref{eq:reformulate1}c) presented at the top of the next page.
Then, constraint (\ref{eq:solve phi org}b) is further re-arranged as
\be\label{eq:new radar constraint}\begin{aligned}
&\Re\big\{ \mathbf{u}^H(\mathbf{I}\otimes\mathbf{h}_{\text{d},\text{t}}\mathbf{h}_{\text{d},\text{t}}^T)\mathbf{w}
+ \mathbf{u}^H\mathbf{F}\bm{\phi} + \mathbf{u}^H\mathbf{L}\text{vec}\{\bm{\phi}\bm{\phi}^T\}\big\}\\
& = \Re\big\{ \mathbf{u}^H(\mathbf{I}\otimes\mathbf{h}_{\text{d},\text{t}}\mathbf{h}_{\text{d},\text{t}}^T)\mathbf{w}
+ \mathbf{u}^H\mathbf{F}\bm{\phi} + \bm{\phi}^T\widetilde{\mathbf{L}}\bm{\phi}\big\} \geq \varepsilon_3,
\end{aligned}\ee
where $\widetilde{\mathbf{L}}\in\mathbb{C}^{N\times N}$ is a reshaped version of $\mathbf{L}^T\mathbf{u}^*$.

Now, it is clear that the third term in (\ref{eq:new radar constraint}) is a complex-valued non-concave function, which leads to an intractable constraint.
To solve this problem, we convert the complex-valued function $-\Re\{\bm{\phi}^T\widetilde{\mathbf{L}}\bm{\phi}\}$ into a real-valued one $\overline{\bm{\phi}}^T\overline{\mathbf{L}}\overline{\bm{\phi}}$ by defining $\overline{\bm{\phi}} \triangleq [\Re\{\bm{\phi}^T\}~\Im\{\bm{\phi}^T\}]^T$ and $\overline{\mathbf{L}}\triangleq \bigg[\begin{array}{cc}
-\Re\{\widetilde{\mathbf{L}}\} & \Im\{\widetilde{\mathbf{L}}\} \\
\Im\{\widetilde{\mathbf{L}}\} & \Re\{\widetilde{\mathbf{L}}\}                                                                                    \end{array}\bigg]$, and then employ the idea of the MM method to seek a series of tractable surrogate functions for it.
In particular, with the solution $\widehat{\bm{\phi}}$ obtained in the previous iteration, an approximate upper-bound for $\overline{\bm{\phi}}^T\overline{\mathbf{L}}\overline{\bm{\phi}}$ is constructed by using the second-order Taylor expansion as\setcounter{equation}{22}
\begin{subequations}\label{eq:Taylor}\begin{align}
\overline{\bm{\phi}}^T\overline{\mathbf{L}}\overline{\bm{\phi}}
&\leq \widehat{\bm{\phi}}^T\overline{\mathbf{L}}\widehat{\bm{\phi}} + \widehat{\bm{\phi}}^T(\overline{\mathbf{L}}+\overline{\mathbf{L}}^T)(\overline{\bm{\phi}}-\widehat{\bm{\phi}}) \non \\
&\qquad\qquad + \frac{\lambda}{2}(\overline{\bm{\phi}}-\widehat{\bm{\phi}})^T(\overline{\bm{\phi}}-\widehat{\bm{\phi}})\\
& = \Re\{\widehat{\bm{\phi}}^T(\overline{\mathbf{L}}\!+\!\overline{\mathbf{L}}^T\!-\!\lambda\mathbf{I})\mathbf{U}\bm{\phi}\} -\widehat{\bm{\phi}}^T\overline{\mathbf{L}}^T\widehat{\bm{\phi}} + \lambda N,
\end{align}\end{subequations}
where $\lambda$ is the maximum eigenvalue of matrix $(\overline{\mathbf{L}}+\overline{\mathbf{L}}^T)$, $\mathbf{U}\triangleq[\mathbf{I}~\jmath\mathbf{I}]$ converts a real-valued expression into a complex-valued one, and $\overline{\bm{\phi}}^T\overline{\bm{\phi}} = \widehat{\bm{\phi}}^T\widehat{\bm{\phi}} = N$ due to the unit modulus property of the reflecting coefficients.
Thus, plugging the result in (\ref{eq:Taylor}) into (\ref{eq:new radar constraint}), the radar output SNR constraint in each iteration can be concisely re-formulated as
\be\label{eq:radar constraint for phi}
\Re\{\widetilde{\mathbf{u}}^H\bm{\phi}\}\leq \varepsilon_4,
\ee
where we define $\widetilde{\mathbf{u}} \triangleq (-\mathbf{u}^H\mathbf{F} +\widehat{\bm{\phi}}^T(\overline{\mathbf{L}}+\overline{\mathbf{L}}^T\!-\!\lambda\mathbf{I})\mathbf{U})^H$ and $\varepsilon_4 \triangleq  -\varepsilon_3 + \widehat{\bm{\phi}}^T\overline{\mathbf{L}}^T\widehat{\bm{\phi}} + \Re\{ \mathbf{u}^H(\mathbf{I}\otimes\mathbf{h}_{\text{d},\text{t}}\mathbf{h}_{\text{d},\text{t}}^T)\mathbf{w}\}-\lambda N$.

Then, we investigate the ADMM method to solve for $\bm{\phi}$ under the unit modulus constraint (\ref{eq:solve phi org}c) as well as the radar constraint derived in (\ref{eq:radar constraint for phi}).
Specifically, an auxiliary variable $\bm{\varphi}\triangleq[\varphi_1,\ldots,\varphi_N]^T$ is introduced to transform the optimization problem of solving for $\bm{\phi}$ into
\begin{subequations}\label{eq:auxiliary phi}\begin{align}
&\underset{\bm{\phi},\bm{\varphi}}\min~~\bm{\phi}^H\mathbf{D}\bm{\phi}-\Re\{\mathbf{g}^H\bm{\phi}\}\\
&~\text{s.t.}~~\Re\{\widetilde{\mathbf{u}}^H\bm{\phi}\}\geq \varepsilon_4,\\
&\quad\quad~|\phi_n| \leq 1,~~\forall n,\\
&\quad\quad~\bm{\phi} = \bm{\varphi},\\
&\quad\quad~|\varphi_n| = 1,~~\forall n.
\end{align}
\end{subequations}
Based on the ADMM method, the solution to (\ref{eq:auxiliary phi}) can be obtained by solving its augmented Lagrangian function:
\begin{subequations}\label{eq:solve for phi and variphi}
\begin{align}
&\underset{\bm{\phi},\bm{\varphi},\bm{\mu}}\min~~\bm{\phi}^H\mathbf{D}\bm{\phi}-\Re\{\mathbf{g}^H\bm{\phi}\} + \frac{\rho}{2}\|\bm{\phi}-\bm{\varphi}+\bm{\mu}/\rho\|^2\\
&~\text{s.t.}\quad~\Re\{\widetilde{\mathbf{u}}^H\bm{\phi}\}\geq \varepsilon_4,\\
&\quad\quad\quad|\phi_n| \leq 1,~~\forall n,\\
&\quad\quad\quad|\varphi_n| = 1,~~\forall n,
\end{align}
\end{subequations}
where $\bm{\mu}\in\mathbb{C}^N$ is the dual variable and $\rho>0$ is a pre-set penalty parameter.
This multi-variate problem can be solved by alternately updating each variable given the others.

\textbf{Update $\bm{\phi}$}:
It is obvious that with fixed $\bm{\varphi}$ and $\bm{\mu}$, the optimization problem for updating $\bm{\phi}$ is convex and can be readily solved by various existing efficient algorithms.

\textbf{Update $\bm{\varphi}$}:
Given $\bm{\phi}$ and $\bm{\mu}$, the optimal $\bm{\varphi}^\star$ can be easily obtained by the phase alignment
\be\label{eq:update varphi}
\bm{\varphi}^\star = e^{\jmath\angle(\rho\bm{\phi}+\bm{\mu})}.
\ee

\textbf{Update $\bm{\mu}$}:
After obtaining $\bm{\phi}$ and $\bm{\varphi}$, the dual variable $\bm{\mu}$ is updated by
\be\label{eq:update mu}
\bm{\mu}:=\bm{\mu} + \rho(\bm{\phi}-\bm{\varphi}).
\ee

\subsection{Summary}

Based on above derivations, the proposed joint beamforming and reflection design algorithm is straightforward and summarized in Algorithm 1.
In the inner loop, we alternately optimize problem (\ref{eq:solve for phi and variphi}) by updating $\bm{\phi}$, $\bm{\varphi}$, and $\bm{\mu}$ to solve for $\bm{\phi}$.
In the outer loop, the auxiliary variables $\mathbf{r}$ and $\mathbf{c}$, the receive filter $\mathbf{u}$, the transmit beamforming $\mathbf{w}$, and the reflection coefficients $\bm{\phi}$ are iteratively updated until convergence.

\begin{algorithm}[!t]
\begin{small}
\caption{Joint Beamforming and Reflection Design}
\label{alg}
    \begin{algorithmic}[1]
    \REQUIRE $\mathbf{h}_{\text{d},\text{t}}$, $\mathbf{h}_{\text{r},\text{t}}$, $\mathbf{G}$, $\sigma_\text{t}^2$, $\sigma_\text{r}^2$, $\mathbf{h}_{\text{d},k}$, $\mathbf{h}_{\text{r},k}$, $\sigma_k^2$, $\forall k$, $P$, $L$, $\Gamma_\text{t}$, $\rho$.
    \ENSURE $\mathbf{W}^\star$, $\bm{\phi}^\star$, and $\mathbf{u}^\star$.
        \STATE {Initialize $\bm{\phi}$ and $\mathbf{W}$ by maximizing channel gains \cite{Liu-2021}.}
        \WHILE {no convergence }
            \STATE{Update $r_k,~\forall k$, by (\ref{eq:update rk}).}
            \STATE{Update $c_k,~\forall k$, by (\ref{eq:update ck}).}
            \STATE{Update $\mathbf{u}$ by (\ref{eq:update u}).}
            \STATE{Update $\mathbf{w}$ by solving problem (\ref{eq:solve W}).}
            \WHILE {no convergence }
            \STATE{Update $\bm{\phi}$ by solving problem (\ref{eq:solve for phi and variphi}) given other variables.}
            \STATE{Update $\bm{\varphi}$ by (\ref{eq:update varphi}).}
            \STATE{Update $\bm{\mu}$ by (\ref{eq:update mu}).}
            \ENDWHILE
        \ENDWHILE
        \STATE{Reshape $\mathbf{w}$ to $\mathbf{W}$.}
        \STATE{Return $\mathbf{W}^\star = \mathbf{W}$, $\bm{\phi}^\star = \bm{\phi}$, and $\mathbf{u}^\star = \mathbf{u}$.}
    \end{algorithmic}
    \end{small}
\end{algorithm}

\section{Simulation Results}

\begin{figure*}[!t]
\centering
\begin{minipage}[t]{0.325\textwidth}\centering\setcaptionwidth{2in}
\includegraphics[width = 2.55in]{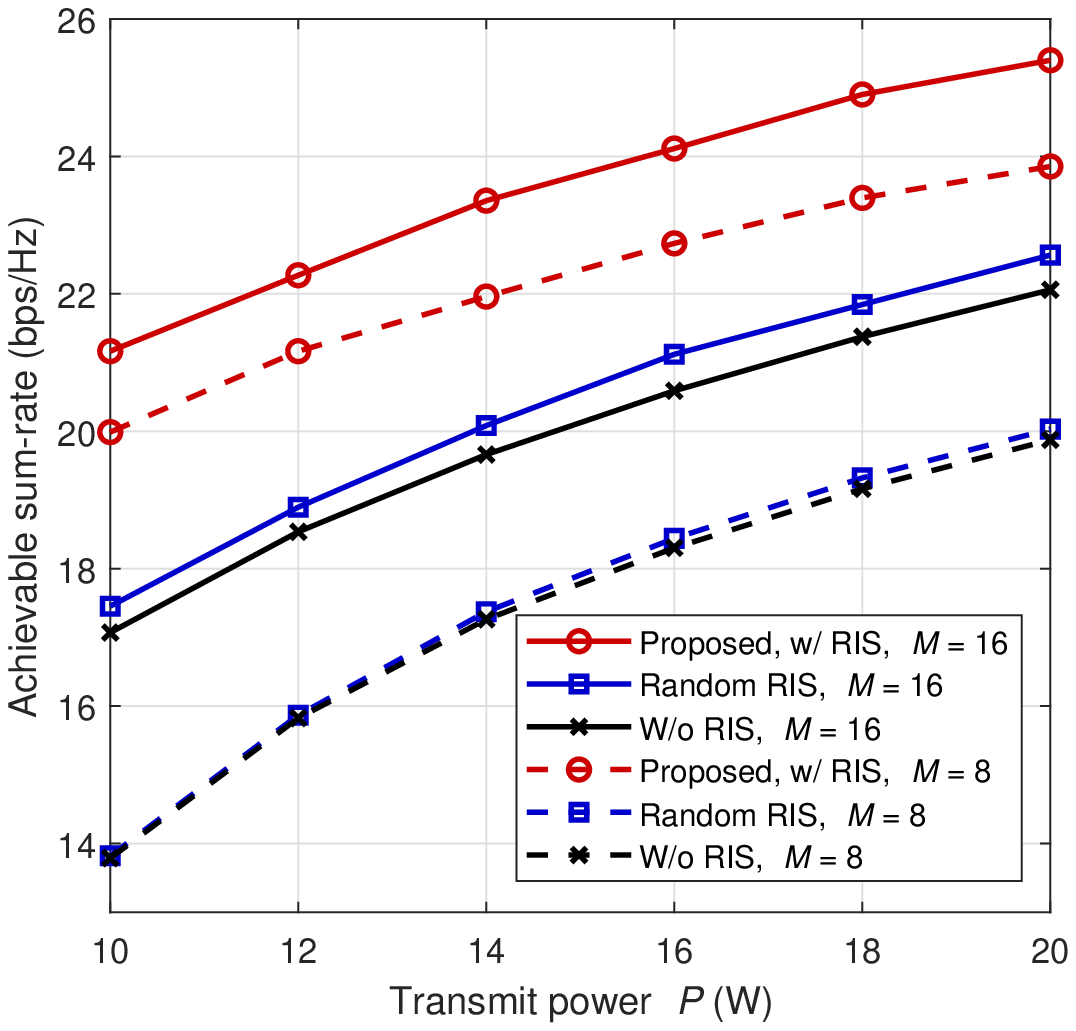}
\caption{Achievable sum-rate versus transmit power $P$ ($N = 100$, $\Gamma_\text{t}=5$dB).}
\end{minipage}
\begin{minipage}[t]{0.325\textwidth}
\includegraphics[width=2.55in]{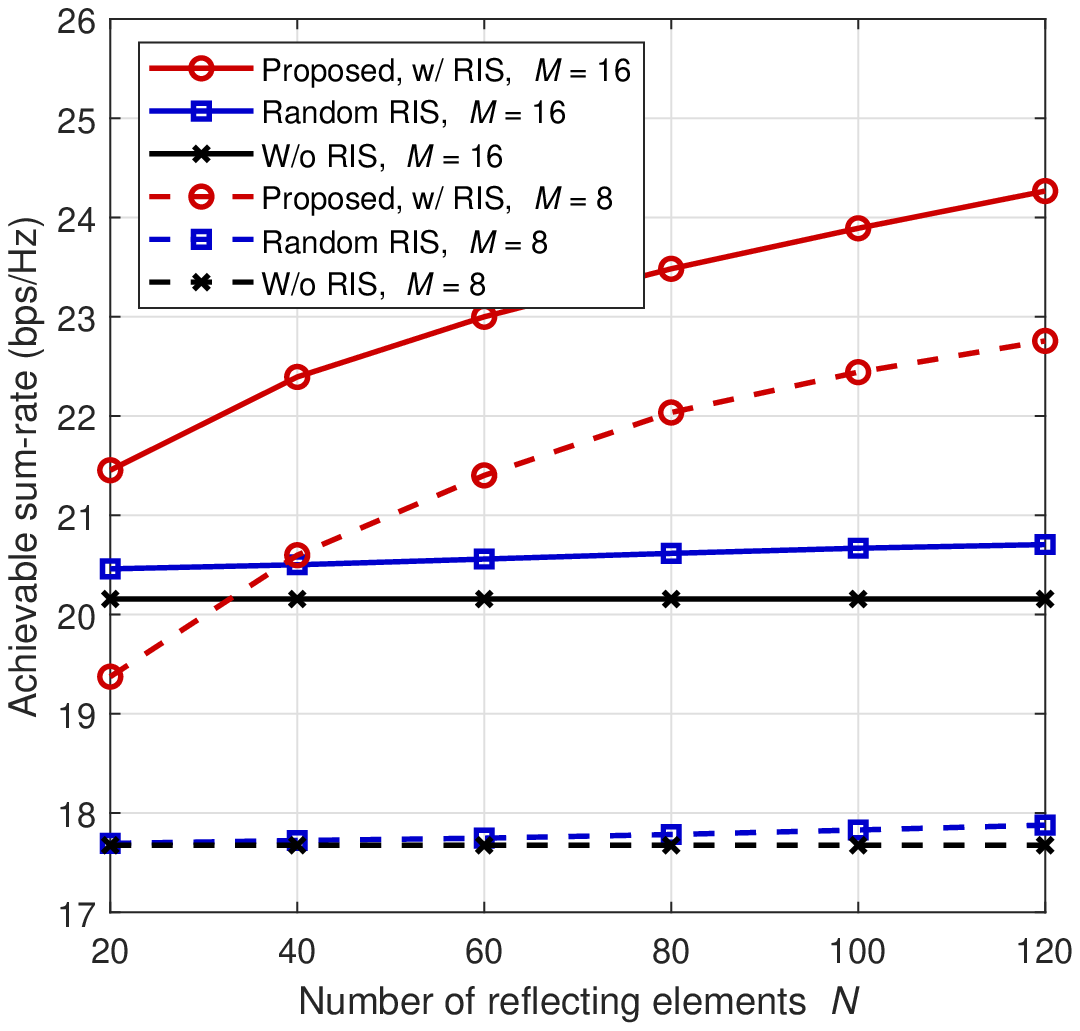}\centering\setcaptionwidth{2in}
\caption{Achievable sum-rate versus the number of reflecting elements $N$ ($P = 15$W, $\Gamma_\text{t} = 5$dB).}
\end{minipage}
\begin{minipage}[t]{0.325\textwidth}\centering\setcaptionwidth{2in}
\includegraphics[width=2.55in]{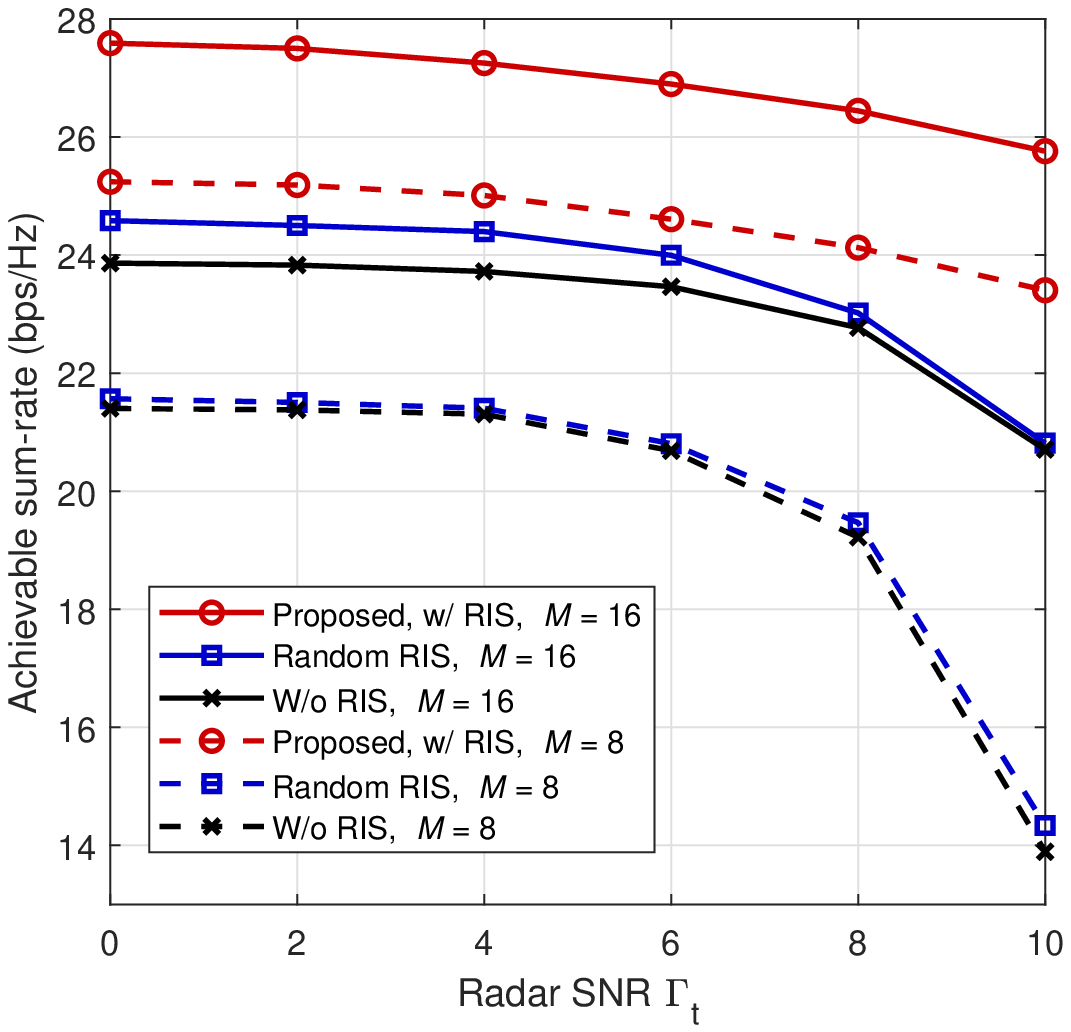}
\caption{Achievable sum-rate versus radar SNR $\Gamma_\text{t}$ ($N = 100$, $P = 25$W).}
\end{minipage}\vspace{0.1 cm}
\end{figure*}

In this section, we present simulation results to verify the advantages of the proposed joint beamforming and reflection design algorithm.
We assume that the BS simultaneously serves $K = 4$ single-antenna users and attempts to detect one potential target at the azimuth angle $0^\circ$ with respect to the BS and the RIS.
The noise power is set as $\sigma_\text{r}^2 = \sigma_k^2 = -80$dBm, $\forall k$, the radar cross section (RCS) is $\sigma_\text{t}^2 = 1$, and the number of collected samples is $L = 1000$.
We adopt a typical distance-dependent path-loss model \cite{Wu-TWC-2019} and set the distances for the BS-RIS, RIS-target, and RIS-user links as 50m, 3m, and 8m, and for the BS-target and BS-user links as $[50,53]$m and $[50,58]$m, respectively.
The path-loss exponents for these links are 2.2, 2.2, 2.3, 2.4, and 3.5, respectively.
Since the users are several meters farther away from the target, the reflected signals from the target to the users are ignored due to severe channel fading.
In addition, the channels of the BS-user and RIS-user links follow the Rayleigh fading model and the others are LoS.

We first present the achievable sum-rate versus the transmit power $P$ in Fig.~1.
In addition to the proposed algorithm (denoted as ``Proposed, w/ RIS''), we also include schemes with random reflecting coefficients (``Random RIS'') and without RIS (``W/o RIS'') for comparison.
We observe that the proposed approach achieves a remarkable performance improvement compared without using an RIS, while using random RIS phases only provides a marginal gain.
In addition, the scenarios with $M=16$ achieve better performance than their counterparts with $M=8$ thanks to more spatial DoFs.
Next, we illustrate the achievable sum-rate versus the number of reflecting elements $N$ in Fig.~2.
It is obvious that more reflecting elements provide larger passive beamforming gain since they exploit more DoFs to manipulate the propagation environment.
Finally, the sum-rate versus the radar SNR $\Gamma_\text{t}$ is shown in Fig.~3, where the trade-off between the performance of multi-user communications and radar target detection can be clearly observed.
These results demonstrate the significant role that RIS can play in improving the performance of ISAC systems and the effectiveness of the proposed algorithm.

\section{Conclusions}

In this paper, we investigated joint beamforming and reflection design for RIS-assisted ISAC systems.
The achievable sum-rate for multi-user communications was maximized under a worst-case radar SNR constraint, the transmit power budget, and the unit modulus restriction on the reflecting coefficients.
An efficient algorithm based on FP, MM, and ADMM methods was developed to convert the resulting non-convex problem into several tractable sub-problems and then iteratively solve them.
Simulation results illustrated the advantages of deploying RIS in ISAC systems and the effectiveness of our proposed algorithm.

\end{document}